\documentclass{emulateapj}
\usepackage{lscape, longtable, hyperref}

\newcommand{\lapprox }{{\lower0.8ex\hbox{$\buildrel <\over\sim$}}}
\newcommand{\gapprox }{{\lower0.8ex\hbox{$\buildrel >\over\sim$}}}
\def\amin{\ifmmode^{\prime}\else$^{\prime}$\fi}
\def\asec{\ifmmode^{\prime\prime}\else$^{\prime\prime}$\fi}

\slugcomment{To appear in the Astrophysical Journal Letters}
\shorttitle{NO NS COMPANION TO J0917$+$4638}
\shortauthors{AG{\" U}EROS ET AL.}

\begin{document}

\title{No Neutron Star Companion To The Lowest Mass SDSS White Dwarf}

\author{
Marcel A. Ag\"ueros\altaffilmark{1,8},
Craig Heinke\altaffilmark{2},
Fernando Camilo\altaffilmark{1},
Mukremin Kilic\altaffilmark{3,9},
Scott F.\ Anderson\altaffilmark{4},
Paulo Freire\altaffilmark{5},
Scot J.\ Kleinman\altaffilmark{6},
James W.\ Liebert\altaffilmark{7},
Nicole M.\ Silvestri\altaffilmark{4}
}

\altaffiltext{1}{Columbia Astrophysics Laboratory, Columbia University, New York, NY 10027; marcel@astro.columbia.edu} 
\altaffiltext{2}{Department of Physics, University of Alberta, Edmonton, AB T6G 2G7, Canada}
\altaffiltext{3}{Harvard-Smithsonian Center for Astrophysics, Cambridge, MA 02138}
\altaffiltext{4}{Department of Astronomy, University of Washington, Seattle, WA 98195}
\altaffiltext{5}{Arecibo Observatory, Arecibo, Puerto Rico 00612}
\altaffiltext{6}{Gemini Observatory, Northern Operations Center, Hilo, HI 96720}
\altaffiltext{7}{Steward Observatory, University of Arizona, Tucson, AZ 85721}
\altaffiltext{8}{NSF Astronomy \& Astrophysics Postdoctoral Fellow}
\altaffiltext{9}{Spitzer Fellow}

\begin{abstract}
SDSS J091709.55$+$463821.8 (hereafter J0917$+$4638) is the lowest surface gravity white dwarf (WD) currently known, with log $g = 5.55 \pm 0.05$ \citep[$M \approx 0.17 M_\odot$;][]{kilic07a, kilic07b}. Such low-mass white dwarfs (LMWDs) are believed to originate in binaries that evolve into WD/WD or WD/neutron star (NS) systems. An optical search for J0917$+$4638's companion showed that it must be a compact object with a mass $\geq 0.28$~M$_\odot$ \citep{kilic07b}. Here we report on Green Bank Telescope $820$ MHz and {\it XMM-Newton} X-ray observations of J0917$+$4638 intended to uncover a potential NS companion to the LMWD. No convincing pulsar signal is detected in our radio data. Our X-ray observation also failed to detect X-ray emission from J0917$+$4638's companion, while we would have detected any of the millisecond radio pulsars in 47 Tuc. We conclude that the companion is almost certainly another WD.
\end{abstract}


\section{Introduction}
Low-mass white dwarfs (LMWDs), generally defined as having $M < 0.45\ M_\odot$, make up a small but highly interesting subset of white dwarfs (WDs). Using the Palomar Green Survey, \citet{liebert05} estimated that the formation rate of LMWDs is $0.4\times10^{-13}$ pc$^{-3}$ yr$^{-1}$, meaning that they make up only $\sim$10$\%$ of the population of the most commonly observed WDs, hydrogen atmosphere DAs. But it is their presumed evolutionary histories that make LMWDs truly intriguing. The youngest WDs in the oldest globular clusters in the Milky Way have masses of $\sim$0.5 $M_\odot$ \citep{hansen07}, implying that lower mass WDs undergo significant mass loss as they form. The preferred scenario is that these WDs form in a tight binary whose evolution includes a phase of mass transfer. As a result, much of the WD progenitor's envelope is removed, preventing a helium flash in its core, and producing a low-mass, helium-core WD.

\citet{brown06} identified J0917$+$4638 as a DA WD in their hyper-velocity star survey of photometrically selected B-star candidates. Detailed model atmosphere analyses by \citet{kilic07a,kilic07b} showed that it has $T_{\rm eff}= 11,855$ K, log $g$ = 5.55, and $M\approx0.17 M_\odot$. The lack of evidence of a companion in the optical photometry forces any main-sequence companion to have $M < 0.1\ M_\odot$, ruling out a low-mass main-sequence star companion. Radial velocity monitoring uncovered variations with a period of 7.6 hr, implying that the mass of the companion is $\geq 0.28\ M_\odot$ \citep{kilic07b}.

What is the nature of this companion? While LMWDs are found in WD/WD systems \citep[e.g.,][]{marsh95}, most known LMWDs are found as companions to neutron stars (NSs), and specifically to NSs ``recycled'' as millisecond pulsars \citep[MSPs;][]{panei07}. Most field radio pulsars in binary systems are MSPs, where a middle-aged, radio-quiet NS has been reactivated as a pulsar via accretion from its companion. The MSP companions are generally thought to be LMWDs with $M = 0.1 - 0.4\ M_\odot$, although they are often too faint for optical spectroscopy to confirm that they are LMWDs \citep[see][]{vankerkwijk05}. Still, a third of the $\sim$50 MSP companions discovered outside of globular clusters have $M\ \lapprox\ 0.2\ M_\odot$, assuming the systems have a median inclination of $60^{\circ}$ \citep[][]{manchester05}. 

While simulations designed to identify the evolutionary pathways that produce LMWD/MSP systems do not generally predict many systems with $P_{\rm orb}$ much shorter than a day \citep[e.g.,][]{nelson04}, and while the system's mass function implies that the probability that J0917$+$4638 has a WD companion is 89\% \citep{kilic07b}, a NS (or black hole) companion to this LMWD cannot be ruled out with the current optical observations. In addition, for the currently known sample of WD/WD systems where both WD masses have been measured, the mass ratio is typically about unity \citep[see][and references therein]{nelemans05}, while the ratio for the J0917$+$4638 binary system is $\leq 0.61$.

\begin{deluxetable*}{ccccccccc}
\tabletypesize{\scriptsize}
\tablecaption{SDSS J0917$+$4638: Properties and Observations\label{props}}
\tablehead{
\colhead{SDSS $g$} & \colhead{$T_{\rm eff}$} & \colhead{$M_{\rm WD}$} & \colhead{$P_{\rm orb}$} & \colhead{Dist.} & \colhead{$b$}        & \colhead{DM}  &\colhead{GBT}  &\colhead{{\it XMM}}\\
\colhead{(mag)}     & \colhead{(K)}           & \colhead{($M_\odot$)}  & \colhead{(hr)}           & \colhead{(kpc)} & \colhead{($^{\circ}$)} & \colhead{(cm$^{-3}$ pc)} & \colhead{Int.\ (s)} & \colhead{Int.\ (s)}
}
\startdata
$18.77\pm0.02$ & $11855$ & $\approx0.17$ & $7.594\pm0.002$ & $2.3$ & $+44.0$ & $80$ & $13300$ & $23418$
\enddata
\tablecomments{The $g$ (PSF) magnitude is from SDSS Data Release 7 \citep{dr7paper}. The distance and orbital period are from \citet{kilic07b}. The listed DM is the maximum value used when searching for pulsations; it corresponds approximately to twice the maximum value obtained in the direction of J0917$+$4638 with the \citet{cordes02} model.} 
\end{deluxetable*}

Because of the connections between LMWDs and MSPs, we used the Green Bank Telescope (GBT) to search for a putative pulsar companion to J0917$+$4638, and report here on these observations (Section~\ref{radio}). We also report on an {\it XMM-Newton X-ray Observatory} observation of this LMWD (Section~\ref{xray}). Blackbody emission from a putative NS companion to the LMWD should be gravitationally bent, allowing us to detect the NS in X~rays even if it were radio-quiet or if its pulsar beam were missing our line of sight \citep{belo02}. We are specifically motivated by the X-ray detection of all known MSPs in the globular cluster 47 Tuc \citep{Heinke05a,Bogdanov06}, allowing predictions of the X-ray emission of other MSPs. We choose this sample of MSPs for comparison in part because the distance to globular clusters such as 47 Tuc \citep[4.5 kpc, 2003 update of][]{harris96} are better known than the distances to most MSPs. We discuss the significance of our nondetections and conclude in Section~\ref{concl}.

\section{Green Bank Telescope Observations}\label{radio}
J0917$+$4638 was observed with the GBT on 2007 November 30. The observing set-up and data reduction were the same as described in \citet{agueros09}. At 820~MHz, the Berkeley-Caltech Pulsar Machine \citep[][]{backer97} provided 48~MHz of bandwidth split into 96 spectral channels; total power samples for each channel were recorded every $72\,\mu$s. The total observing time was $13,300$ s ($3.7$ hr). We used standard pulsar search techniques as implemented in the PRESTO software package \citep{ransom01}. We calculated the maximum dispersion measure (DM) expected in the direction of J0917$+$4638 using the \citet{cordes02} model for the distribution of free electrons in the Galaxy. To account for uncertainties, we dedispersed the data up to a DM limit twice that obtained from the model, i.e., DM $= 80$ cm$^{-3}$ pc.

No convincing pulsar signal was detected in our data. Below we discuss the limitations of our search.

\subsection{Acceleration Sensitivity}
The orbital motion of a putative pulsar companion to J0917$+$4638 could significantly affect its apparent spin period. Based on radial velocity monitoring, \citet{kilic07b} found that J0917$+$4638 is in an orbit with a period $7.6$ hr. Assuming that the LMWD companion is a $1.4\ M_\odot$ NS, this implies that the maximum orbital acceleration is on the order of $100$~m~s$^{-2}$, which is significantly larger than what is typically seen in these systems \citep[for $90\%$ of known pulsars the maximum orbital acceleration is $\leq |25|$ m s$^{-2}$;][]{manchester05, joeri07}. 

Our integration time represents nearly half of the binary orbital period. As a result, the assumption of a constant apparent acceleration built into PRESTO breaks down. We therefore divided our GBT data into $14$ separate $900$~s integrations (each representing $\sim3\%$ of an orbit) and one $700$~s integration and conducted searches for pulsations separately in each of these partial observations.\footnote{We also conducted a search of the entire $3.7$~hr integration, which unsurprisingly did not return any good candidate pulsar signals.} This extended our search sensitivity to accelerations on the order of several hundred m~s$^{-2}$, but as detailed in the following section, reduced our luminosity sensitivity. None of these searches uncovered a convincing pulsar signal.

\subsection{Luminosity Sensitivity}\label{lum_sens}
We use the standard modifications to the radiometer equation to calculate the minimum detectable period-averaged flux density for our searches. We consider a pulsar duty cycle of $20\%$ (typical of MSPs). At $820$~MHz, the GBT gain is 2\,K\,Jy$^{-1}$ and the system temperature is 25\,K. The sky temperature at this frequency and a Galactic latitude of $b =+44^{\circ}$ only adds a few K to the overall temperature. We consider an effective threshold signal-to-noise ratio of 10. For $t_{int} = 900$~s, the sensitivity limit for a long period pulsar at the beam center is $\sim$0.26 mJy. 

Pulsar luminosities are often measured at 1400\,MHz; using a typical spectral index of $-1.7$, the limiting sensitivity at that frequency is $S_{1400} \approx 0.10$~mJy when searching the $900$~s integrations. For an MSP period of 3~ms, our sensitivity at 1400~MHz was roughly $0.14$~mJy for each integration, and it quickly degraded for shorter periods; it was $10\times$ worse for 1~ms. 

The distance to J0917$+$4638 is estimated to be $2.3$ kpc \citep{kilic07b}, implying that our $L_{1400} \equiv S_{1400} d^2$ limits for 3\,ms periods are $\approx\ 0.7$ mJy~kpc$^2$ for each $900$~s integration. According to the ATNF's pulsar catalog\footnote{{\tt http://www.atnf.csiro.au/research/pulsar/psrcat/}.} \citep{atnf}, of the $50$ MSPs (periods $<25$ ms) outside of globular clusters and with measured luminosities, $64\%$ have $L_{1400} > 0.7$ mJy~kpc$^2$. We would therefore expect our search to detect roughly two-thirds of the known MSPs were one orbiting J0917$+$4638 and beaming radio waves toward the Earth. 

We note that J0917$+$4638 falls within the FIRST footprint and is not detected in that $1.4$ GHz survey, for which the sensitivity limit is roughly $1$~mJy \citep{first}.

\section{XMM-Newton Observation}\label{xray}
\subsection{Motivation}
MSP radio beaming fractions are $< 100\%$, and as a result, some MSPs have not yet been detected in the radio in binary systems where there is strong evidence for their presence \citep[e.g., the companion to the young pulsar PSR J1906$+$0746;][]{lorimer06}. Given that the NS blackbody emission is gravitationally bent, allowing us to view $>75\%$ of the NS surfaces in X rays \citep{belo02}, sufficiently deep X-ray observations are virtually guaranteed to detect these MSPs.  

\citet{Heinke05a} found no correlation between the X-ray and radio luminosities of MSPs in 47 Tuc, as expected due to the differing nature and spatial location of the X-ray and radio emission, and found that all MSPs in 47 Tuc\footnote{Only 15 MSPs in 47 Tuc have published locations farther than $1\asec$ from other MSPs; two pairs of MSPs that are closer cannot be conclusively resolved \citep{Bogdanov06}.  However, the flux from each pair is consistent with expectations from the other MSPs.} have X-ray luminosities ranging between $L_X (0.5-6$ keV$) =2\times10^{30}$ and $2\times10^{31}$ erg s$^{-1}$.  \citet{Bogdanov06} showed that the X-ray spectra of the MSPs in 47 Tuc are typically dominated by thermal blackbody-like emission from the NS surface around the polar caps, with temperature $1-3\times10^6$ K. This X-ray emission is sometimes overwhelmed by additional non-thermal X rays that are either magnetospheric or due to an intra-binary shock. \citet{Bogdanov06} also showed that there are no clear systematic differences between the X-ray properties of MSPs in 47 Tuc and in the Galactic field.  Thus, we requested an {\it XMM} observation capable of detecting any of the known MSPs in 47 Tuc, were they located at the distance of J0917$+$4638. 

J0917$+$4638 had not previously been observed in the X-ray since the {\it ROSAT} All-Sky Survey \citep{voges99,fsc}, where it was not detected (unsurprisingly, considering the short exposure time).

\subsection{X-Ray Data Analysis}
We observed J0917$+$4638 on 2008 May 7 for 17 ks (ObsID 0553440101) with {\it XMM}'s EPIC camera, consisting of two MOS CCD detectors \citep{turner01} and a pn CCD detector \citep{struder01}.  All data were reduced using FTOOLS\footnote{{\tt http://heasarc.gsfc.nasa.gov/docs/software/ftools/ftools\_menu.html}.} and SAS version 8.0.0.\footnote{{\tt http://xmm.vilspa.esa.es}.}  We excluded times of soft proton background flaring, when the pn camera's count rate exceeded 25 ($0.2-10$ keV) counts s$^{-1}$, or when the MOS cameras exceeded 7 or 8 ($0.2-10$ keV) counts s$^{-1}$ for the MOS1 or MOS2 cameras respectively.  This left 8.9 ks of good data from the pn detectors, and 11.3 ks from the MOS detectors.  We filtered the events on pixel patterns (trying PATTERN$<=$1 or $<=$4 for pn and PATTERN$<=$12 for MOS data), and for FLAG==0.  We choose an energy range of $0.2-1.5$ keV to obtain optimal sensitivity to the soft blackbody emission expected from MSPs. 

No source is detected at or within $1\amin$ of the location of J0917$+$4638 either with detection algorithms or by eye.  We utilize our knowledge of the {\it XMM} point spread function\footnote{{\it XMM-Newton} User's Handbook, {\tt http://xmm.esac.esa.int/external/xmm\_user\_support/documentation/uhb/index.html}.} and absolute pointing accuracy \citep[$<1\asec$;][]{Kirsch04} to determine an upper limit.  For the pn, $50\%$ of 1.5 keV photons are found within $8\asec$, and $80\%$ within $20\asec$.  For the MOS cameras, $50\%$ of 1.5 keV photons are recorded within $8\asec$, and $75\%$ within $20\asec$.  We find 3 counts within an $8\asec$ circle, or 20 counts within a $20\asec$ circle, in the combined image.  This is consistent with a nondetection, as the expected background counts in these circles are $3.2\pm0.2$ and $19.8\pm1.3$ counts, respectively, as derived from nearby background regions.  

\subsection{Comparison to MSPs in 47 Tuc}
We use the X-ray faintest MSP in 47 Tuc, 47 Tuc T, to calibrate our expectations for the detection of an MSP in J0917$+$4638.  47 Tuc T has $L_X (0.2-1.5$ keV$)=1.5\times10^{30}$ erg s$^{-1}$ and a 134 eV blackbody spectrum \citep{Bogdanov06}.  We use PIMMS\footnote{{\tt http://asc.harvard.edu/toolkit/pimms.jsp}.} to determine the expected EPIC count rates from 47 Tuc T were it located at 2.3 kpc (the distance to J0917$+$4638) behind an estimated $N_H=1.5\times10^{20}$ cm$^{-2}$ \citep{dickey90}.  We expect 10.9 counts within $8\asec$, or 17.0 within $20\asec$, accounting for the relevant encircled energy fractions, from such an MSP.  Comparing the predicted counts with the Poisson errors on the detected counts \citep[][equation 7]{Gehrels86}, we find that the number of counts within $20\asec$ is $3.0\sigma$ below expectations for the faintest known MSP in 47 Tuc, while the counts within $8\asec$ are $3.5\sigma$ below those expectations.  47 Tuc T is the X-ray faintest of the 15 independently measured MSPs in 47 Tuc; the median X-ray luminosity is $2.1\times$ greater \citep{Bogdanov06}, which is ruled out at $6.3\sigma$ confidence. Our nondetection is therefore strong evidence against the existence of an MSP in J0917$+$4638.

\section{Conclusion}\label{concl}
We have searched for evidence of an MSP companion to the LMWD J0917$+$4638 through radio and X-ray observations. Our radio search reaches a sensitivity sufficient to detect roughly two-thirds of the known MSPs, while our X-ray search is sensitive enough to detect any of the 15 independently identified MSPs in 47 Tuc. Together, our nondetections provide strong evidence against the presence of an MSP in this system. Furthermore, since any NS companion to J0917$+$4638 would presumably have been recycled through accretion from the LMWD, we rule out the presence of a NS in this system. Although a black hole companion is still conceivable (as the 7.6 hr orbital period would not induce current accretion and X-ray activity), such a companion is far less probable than a WD companion given both the system's mass function \citep{kilic07b} and the stellar initial mass function for $M \geq 1 \ M_\odot$ \citep[e.g.,][]{scalo98}. We conclude that J0917$+$4638's more massive companion ($M \geq 0.28\ M_\odot$) is almost certainly another WD.  

Roughly two dozen WD/WD binaries are known and in ten such systems both WD masses have been measured \citep[see][and references therein]{nelemans05}. The individual masses of WDs in these systems range between $0.29$ and $0.71\ M_\odot$; the median mass for those with measured masses (and not just lower limits) is $0.43\ M_\odot$. The majority of these double WD systems have mass ratios near unity, which is contrary to what is expected from standard population synthesis models \citep{nelemans05b}. This has been used to argue that energy balance ($\alpha$-formalism), the standard prescription for common envelope evolution, should be replaced by angular momentum balance \citep[$\gamma$-algorithm;][]{maxted02,nelemans05}.

In particular, \citet{nelemans05} found that the $\alpha$-formalism cannot be used to describe the first phase of mass transfer for nine of the ten double WD systems in which both WD masses have been measured. The exception is WD1704$+$481, which has a mass ratio $=0.7$, similar to the expected ratio from the $\alpha$-formalism. The mass ratio for the J0917$+$4638 binary system is $\leq 0.61$. Recent observations of another LMWD, LP400$-$22, showed 

\clearpage
\noindent that it is in a binary with a mass ratio $\leq 0.46$ \citep{kilic09}. The mass ratios of these systems imply that the $\alpha$-formalism may explain at least some of the WD/WD binaries. Determining the mass ratios of the other SDSS LMWD systems for which the nature of the companion is currently unknown will be important in understanding the role of energy versus momentum balance in reconstructing common envelop evolution.

\section*{Acknowledgments}
M.A.A.\ is supported by an NSF Astronomy and Astrophysics Postdoctoral Fellowship under award AST-0602099. C.H.\ is supported by NSERC. Further support was provided to M.K.\ through the Spitzer Space Telescope Fellowship Program, through a contract issued by the JPL/Caltech under a contract with NASA.  

The Robert C.\ Byrd Green Bank Telescope is operated by the National Radio Astronomy Observatory, which is a facility of the US National Science Foundation operated under cooperative agreement by Associated Universities, Inc. {\it XMM-Newton} is an ESA science mission with instruments and contributions directly funded by ESA Member States and NASA.

\end{document}